%% file: main.tex
\documentclass[a4paper,10pt,twocolumn]{article}
\usepackage[utf8]{inputenc}
\usepackage[left=1.4cm,right=1.4cm, top=1.8cm, bottom=1.8cm]{geometry}
\usepackage{amsmath}%
\usepackage{amssymb}%
\usepackage{bm}%
\usepackage{graphicx}%
\usepackage{pgfplots}%
\usepackage{subcaption}%
\usepackage[nolist]{acronym}%
\usepackage{multicol}%
\usepackage{microtype}%

\title{Data-Driven Uncertainty Modeling for Robust Feedback Active Noise Control in Headphones}%
\author{Florian Hilgemann, Egke Chatzimoustafa, and Peter Jax}%
\date{}%
\input{macros.tex}%
\input{operators.tex}%
\input{plotdef.tex}%
\input{variables.tex}
\begin{document}
\input{acronyms.tex}

\begin{titlepage}
   \begin{center}
       \vspace*{6cm}
       \large
       This draft was submitted to the Journal of the Audio Engineering Society on Aug. 8, 2024. The bibliographic information of the current published version is:

       \vspace{2cm}

       \begin{tabular}{rl}
            Authors: & \hspace{1.2ex}Florian Hilgemann, Egke Chatzimoustafa, and Peter Jax \\
            Title: & \begin{tabular}{l}Data-Driven Uncertainty Modeling for Robust Feedback Active \\ Noise Control in Headphones \end{tabular} \\
            Journal: & \hspace{1.2ex}Journal of the Audio Engineering Society \\
            Volume: & \hspace{1.2ex}72 \\
            Number: & \hspace{1.2ex}12 \\
            Pages: & \hspace{1.2ex}873-883 \\
            Year: & \hspace{1.2ex}2024 \\
            DOI: & \hspace{1.2ex}http://dx.doi.org/10.17743/jaes.2022.0185
       \end{tabular}

       \vfill
       Please cite the article as follows: \vspace{.25cm}

       F. Hilgemann, E. Chatzimoustafa, and P. Jax - ``Data-Driven Uncertainty Modeling for Robust Feedback Active Noise Control'', J. Audio Eng. Soc., vol. 72, no. 12, 2024, pp. 873-883

   \end{center}
\end{titlepage}

\maketitle

\begin{abstract}
\Ac{ANC} has become popular for reducing noise and thus enhancing user comfort in headphones. While feedback control offers an effective way to implement \ac{ANC}, it is restricted by uncertainty of the controlled system that arises, \eg, from differing wearing situations. Widely used unstructured models which capture these variations tend to overestimate the uncertainty and thus restrict \ac{ANC} performance. As a remedy, this work explores uncertainty models that provide a more accurate fit to the observed variations in order to improve \ac{ANC} performance for over-ear and in-ear headphones. We describe the controller optimization based on these models and implement an \ac{ANC} prototype to compare the performances associated with conventional and proposed modeling approaches. Extensive measurements with human wearers confirm the robustness and indicate a performance improvement over conventional methods. The results allow to safely increase the active attenuation of \ac{ANC} headphones by several decibels.
\end{abstract}

\input{introduction.tex}
\input{feedbackcontrol.tex}
\input{headphonemeasurements.tex}
\input{uncertaintymodels.tex}
\input{controlleroptimization.tex}
\input{case_study.tex}
\input{conclusion.tex}
\section{Acknowledgement}

The authors wish to thank students, colleagues, and friends for participating in the measurements.

\bibliographystyle{ieeetr}
\bibliography{literature.bib}

\end{document}

%% file: macros.tex
\newcommand{\eg}{e.g.}%
\newcommand{\ie}{i.e.}%

%% file: operators.tex
\newcommand{\fd}[1]{\MakeUppercase{#1}}

\renewcommand{\vec}[1]{\mathbf{#1}}%
\newcommand{\trans}[1]{#1^{\mathrm{T}}}%
\newcommand{\set}[1]{\left\{ #1 \right\}}%
\newcommand{\abs}[1]{\left| #1 \right|}%
\newcommand{\real}[1]{\Re\left(#1\right)}%
\newcommand{\imag}[1]{\Im\left(#1\right)}%
\newcommand{\conj}[1]{#1 ^ *}%
\newcommand{\average}[1]{\overline{#1}}%

\newcommand{\nominal}[1]{\widehat{#1}}%

\newcommand{\normbounded}[1]{#1^{(\mathrm{NB})}}%
\newcommand{\multidisk}[1]{#1^{(\mathrm{MD})}}%
\newcommand{\elliptic}[1]{#1^{(\mathrm{E})}}%
\newcommand{\convexhull}[1]{#1^{(\mathrm{CH})}}%

\newcommand{\modified}[1]{#1'}%

\newcommand{\figref}[1]{Fig.~\ref{#1}}%
\newcommand{\secref}[1]{Sec.~\ref{#1}}%

%% file: plotdef.tex
\tikzstyle{box}=[draw,rectangle,minimum height=0.65cm,minimum width=1.1cm,rounded corners=1pt,fill=white]%
\tikzstyle{smallbox}=[draw,rectangle,minimum height=0.55cm,minimum width=0.55cm,rounded corners=1pt,fill=white]%
\tikzstyle{dot}=[fill,shape=circle,minimum size=2pt,inner sep=0pt]%
\tikzstyle{operator}=[draw,fill=white,circle, inner sep=0.3pt, font=\scriptsize]%

\newcommand{\refnormbounded}[1][black]{\hspace{-.5ex}%
	\begin{tikzpicture}%
		\node at (0, 0) (dummy1) {};%
		\node at (0.5, 0) (dummy2) {};%
		\draw [color=color_norm_bounded, thin] (dummy1.center) -- (dummy2.center);%
	\end{tikzpicture}\hspace{-.5ex}%
}%
\newcommand{\refcenteredmultidisk}[1][black]{\hspace{-.5ex}%
	\begin{tikzpicture}%
		\node at (0, 0) (dummy1) {};%
		\node at (0.5, 0) (dummy2) {};%
		\draw [color=color_centered_multi_disk, thin] (dummy1.center) -- (dummy2.center);%
	\end{tikzpicture}\hspace{-.5ex}%
}%
\newcommand{\refelliptic}[1][black]{\hspace{-.5ex}%
	\begin{tikzpicture}%
		\node at (0, 0) (dummy1) {};%
		\node at (0.5, 0) (dummy2) {};%
		\draw [color=color_elliptic, thin] (dummy1.center) -- (dummy2.center);%
	\end{tikzpicture}\hspace{-.5ex}%
}%
\newcommand{\refconvexhull}[1][black]{\hspace{-.5ex}%
	\begin{tikzpicture}%
		\node at (0, 0) (dummy1) {};%
		\node at (0.5, 0) (dummy2) {};%
		\draw [color=color_convex_hull, thin] (dummy1.center) -- (dummy2.center);%
	\end{tikzpicture}\hspace{-.5ex}%
}%
\newcommand{\reftarget}[1][black]{\hspace{-.5ex}%
\begin{tikzpicture}%
	\node at (0, 0) (dummy1) {};%
	\node at (0.5, 0) (dummy2) {};%
	\draw [color=black, semithick, densely dashed] (dummy1.center) -- (dummy2.center);%
\end{tikzpicture}\hspace{-.5ex}%
}%
\newcommand{\refdhloose}[1][black]{\hspace{-.5ex}%
	\begin{tikzpicture}%
		\node at (0, 0) (dummy1) {};%
		\node at (0.5, 0) (dummy2) {};%
		\draw [color=color_dummy_head, densely dotted, semithick] (dummy1.center) -- (dummy2.center);%
	\end{tikzpicture}\hspace{-.5ex}%
}%
\newcommand{\refdhnormal}[1][black]{\hspace{-.5ex}%
	\begin{tikzpicture}%
		\node at (0, 0) (dummy1) {};%
		\node at (0.5, 0) (dummy2) {};%
		\draw [color=color_dummy_head, semithick] (dummy1.center) -- (dummy2.center);%
	\end{tikzpicture}\hspace{-.5ex}%
}%
\newcommand{\refdhtight}[1][black]{\hspace{-.5ex}%
	\begin{tikzpicture}%
		\node at (0, 0) (dummy1) {};%
		\node at (0.5, 0) (dummy2) {};%
		\draw [color=color_dummy_head, densely dashed, semithick] (dummy1.center) -- (dummy2.center);%
	\end{tikzpicture}\hspace{-.5ex}%
}%
\newcommand{\refloosefit}[1][black]{\hspace{-.5ex}%
	\begin{tikzpicture}%
		\node at (0, 0) (dummy1) {};%
		\node at (0.5, 0) (dummy2) {};%
		\draw [color=color_loose_fit, thin] (dummy1.center) -- (dummy2.center);%
	\end{tikzpicture}\hspace{-.5ex}%
}%
\newcommand{\refnormalfit}[1][black]{\hspace{-.5ex}%
	\begin{tikzpicture}%
		\node at (0, 0) (dummy1) {};%
		\node at (0.5, 0) (dummy2) {};%
		\draw [color=color_normal_fit, thin] (dummy1.center) -- (dummy2.center);%
	\end{tikzpicture}\hspace{-.5ex}%
}%
\newcommand{\reftightfit}[1][black]{\hspace{-.5ex}%
	\begin{tikzpicture}%
		\node at (0, 0) (dummy1) {};%
		\node at (0.5, 0) (dummy2) {};%
		\draw [color=color_tight_fit, thin] (dummy1.center) -- (dummy2.center);%
	\end{tikzpicture}\hspace{-.5ex}%
}%
\newcommand{\refscatterloose}[1][black]{\hspace{.2ex}%
	\begin{tikzpicture}%
		\node [draw=color_loose_fit, circle, inner sep=2pt] {};%
	\end{tikzpicture}\hspace{.2ex}%
}%
\newcommand{\refscatternormal}[1][black]{%
\textcolor{color_normal_fit}{$\times$}\hspace{-1ex}
}%
\newcommand{\refscattertight}[1][black]{\hspace{.2ex}%
\begin{tikzpicture}%
	\node [draw=color_tight_fit, rectangle, inner sep=2.5pt] {};%
\end{tikzpicture}\hspace{.2ex}%
}%

\definecolor{color_normal_fit}{RGB}{0, 84, 159}%
\definecolor{color_loose_fit}{RGB}{250, 190, 80}%
\definecolor{color_tight_fit}{RGB}{87, 171, 39}%
\definecolor{color_dummy_head}{RGB}{0, 0, 0}%
\definecolor{color_area_background}{RGB}{225, 225, 225}%
\definecolor{color_norm_bounded}{RGB}{189, 205, 0}%
\definecolor{color_norm_bounded_individual}{RGB}{189, 205, 0}%
\definecolor{color_centered_multi_disk}{RGB}{155, 145, 193}%
\definecolor{color_centered_multi_disk_individual}{RGB}{155, 145, 193}%
\definecolor{color_elliptic}{RGB}{0, 152, 161}%
\definecolor{color_elliptic_individual}{RGB}{0, 152, 161}%
\definecolor{color_convex_hull}{RGB}{97, 33, 88}%
\definecolor{color_convex_hull_individual}{RGB}{97, 33, 88}%
\definecolor{color_target}{RGB}{0, 0, 0}%

%% file: variables.tex
\newcommand{\iu}{\mathrm{j}}% imaginary unit
\newcommand{\z}{z}%
\newcommand{\dt}{n}% discrete time

\newcommand{\samplerate}{f_{\mathrm{s}}}%
\newcommand{\frequency}{f}%
\newcommand{\normfreq}{\Omega}%
\newcommand{\normfreqbin}{\Omega_{\freqidx}}%
\newcommand{\onunitcircle}{e^{\iu\normfreq}}%
\newcommand{\onunitcirclebin}{e^{\iu\normfreqbin}}%

\newcommand{\numobservations}{N_{\fd{\plant}}}%
\newcommand{\plantset}{\Pi}%
\newcommand{\weight}{w}%
\newcommand{\circlecenterdiscretefreq}[1]{\fd{\plant}^{(0)}_{#1\freqidx}}%
\newcommand{\muluncdiscretefreq}[1]{\fd{\weight}_{\mathrm{M}#1,\freqidx}}%
\newcommand{\circleradiusdiscretefreq}{R_{\freqidx}}%
\newcommand{\circlesradiusdiscretefreq}[1]{R_{#1\freqidx}}%
\newcommand{\numdisks}{p_{\freqidx}}%
\newcommand{\ellipmajordiscretefreq}{R_{x, \freqidx}}%
\newcommand{\ellipminordiscretefreq}{R_{y, \freqidx}}%
\newcommand{\ellipanglediscretefreq}{\theta_{\freqidx}}%
\newcommand{\ellipcenterdiscretefreq}{\fd{\plant}^{(0)}_{\freqidx}}%
\newcommand{\modifiedellipcenterdiscretefreq}{\fd{\plant}^{(0)'}_{\freqidx}}%
\newcommand{\ellipmajorauxdiscretefreq}{X_{\freqidx}}%
\newcommand{\ellipminorauxdiscretefreq}{Y_{\freqidx}}%
\newcommand{\firsthalfspacevardiscretefreq}{A_{0\halfspaceidx,\freqidx}}%
\newcommand{\secondhalfspacevardiscretefreq}{A_{1\halfspaceidx,\freqidx}}%
\newcommand{\halfspaceoffsetdiscretefreq}{B_{\halfspaceidx,\freqidx}}%
\newcommand{\numhalfspaces}{m_{\freqidx}}%

\newcommand{\disturbance}{d}%
\newcommand{\error}{e}%
\newcommand{\controlsignal}{c}%
\newcommand{\controller}{k}%
\newcommand{\plant}{g}%
\newcommand{\q}{q}%

\newcommand{\sensitivity}{s}%
\newcommand{\numfreq}{N_{\normfreq}}%

\newcommand{\loss}{J}%
\newcommand{\filtercoefficients}{\vec{\q}}%
\newcommand{\numtaps}{N_{\q}}%
\newcommand{\zvec}{\vec{z}}
\newcommand{\performanceweight}{\fd{\weight}_{1}}%
\newcommand{\openloop}{l}

\newcommand{\freqidx}{\mu}%
\newcommand{\observationidx}{i}%
\newcommand{\circleidx}{l}%
\newcommand{\halfspaceidx}{l}%

\newcommand{\convexhullanglediscretefreq}[1]{\alpha_{#1, \freqidx}}%

\newcommand{\performanceweightdiscretefreq}{\fd{\weight}_{1, \freqidx}}%
\newcommand{\plantdiscretefreq}{\fd{\plant}_\freqidx}%
\newcommand{\plantobservationdiscretefreq}{\fd{\plant}_{\observationidx,\freqidx}}%
\newcommand{\openloopdiscretefreq}{\fd{\openloop}_\freqidx}%
\newcommand{\qdiscretefreq}{\fd{q}_\freqidx}%
\newcommand{\controllerdiscretefreq}{\fd{\controller}_\freqidx}%
\newcommand{\plantsetdiscretefreq}{\Pi_{\freqidx}}%

\newcommand{\constraint}{C}%
\newcommand{\constraintdiscretefreq}{\constraint_{\freqidx}}%

\newcommand{\ellipaux}{U_\freqidx}%
\newcommand{\convexhullauxfirst}[1]{V_{#1, \freqidx}}%
\newcommand{\convexhullauxsecond}{W_{\freqidx}}%
\newcommand{\minapproxvar}{\rho}%

%% file: acronyms.tex
\begin{acronym}%
    \acro{ADC}{analog-digital converter}%
    \acro{ANC}{active noise control}%
    \acro{AOC}{active occlusion cancellation}%
    \acro{DAC}{digital-analog converter}%
    \acro{DFT}{discrete Fourier transform}%
	\acro{DSP}{digital signal processor}%
    \acro{FIR}{finite impulse response}%
    \acro{IIR}{infinite impulse response}%
    \acro{IMC}{internal model control}%
    \acro{LTI}{linear time-invariant}%
    \acro{PSD}{power spectral density}%
    \acro{SNR}{signal-to-noise ratio}%
    \acro{SQP}{sequential quadratic programming}%
\end{acronym}%

%% file: introduction.tex
\section{Introduction}%
\label{sec:introduction}%

Since the initial patent by Lueg~\cite{lueg36} and early works by Olson and May~\cite{olson53}, \ac{ANC} has become a ubiquitous feature in hearing devices such as headphones or hearables~\cite{gupta22}. It was implemented using analog circuitry in the past, but is nowadays based on fixed or adaptive digital filters~\cite{elliott93,hansen12}. Its popularity has risen due to the provided cost-efficient reduction of ambient noise, especially for low frequencies.

A widely-used variant to implement \ac{ANC} is feedback control. It bears a hazard of instability, which requires proper consideration in the design. In contrast to feedforward control, feedback control does not require an external microphone to measure the noise source. Feedback \ac{ANC} is applied beyond noise reduction purposes to improve the speech perception of headphone wearers by mitigating the occlusion effect~\cite{mejia08, denk24}. 

This work focuses on the design of feedback controllers for \ac{ANC} headphones. A fundamental challenge is that the controlled system changes, \eg, with the headphone fit on or in the wearer's ears. Classical measures to ensure closed-loop stability, such as gain or phase margins, have been adopted in~\cite{yu01, benois22}. However, such margins cannot model simultaneous gain and phase variations in the controlled system~\cite{seiler20}. This work focuses on methods that employ the more rigorous robust stability specification, as studied by Zames~\cite{zames81}. These methods rely on a model of the controlled system's variations. This so-called uncertainty is commonly modeled using disks in the complex plane. State-of-the-art controller design methods that guarantee robust stability are based on $\mathcal{H}_{\infty}$-synthesis~\cite{bai97} or constrained least-squares optimization~\cite{rafaely99, wang21}.

Studies have shown that variations in \ac{ANC} systems occur with frequency-dependent shapes that are often not modeled accurately by means of disks~\cite{rafaely99, hilgemann22, guo24}. In these situations, disk-based models overestimate the uncertainty, which implies suboptimal \ac{ANC} performance due to overly conservative design constraints. Uncertainty models that better reflect the variations have been described in~\cite{east82, laughlin86} as part of a computer-aided manual controller tuning process, but were deemed too complicated for controller design~\cite{skogestad05}, and not considered in \ac{ANC} applications for a long time. Recent works that aim to reduce the modeling conservatism to improve \ac{ANC} performance include our previous work that uses multiple, smaller disks~\cite{hilgemann22}, or another approach based on rectangular shapes~\cite{guo24}.

This work proposes uncertainty models for the improved design of feedback \ac{ANC} controllers. The main difference to other approaches is a more accurate uncertainty description that we use for the optimization. After a brief technical overview in~\secref{sec:feedback_control_in_headphones}, we analyze frequency response variations in \ac{ANC} headphones to motivate the use of such models in~\secref{sec:headphone_measurements}. We define the proposed uncertainty models in~\secref{sec:uncertainty_modeling}, and integrate them into an existing controller design method in~\secref{sec:controller_optimization}. Finally, \secref{sec:case_study} details the design and implementation of a real-time \ac{ANC} prototype based on the conventional and the proposed models. It compares their robustness and performance based on measurements with human wearers and a dummy head.

%% file: feedbackcontrol.tex
\section{Feedback Control in Headphones}%
\label{sec:feedback_control_in_headphones}%

In this section, we revisit the mode of operation for \ac{ANC} headphones using standard notation: we write digital signals as a function of the discrete time index $\dt$ with sampling rate $\samplerate$. Upper-case letters indicate frequency domain variables, \eg, $\fd{\error}(\z)$ and $\fd{\error}(\onunitcircle)$ refer to the $\z$-transform and Fourier transform of $\error(\dt)$, respectively, and $\normfreq = \frac{2\pi\frequency}{\samplerate}$ denotes the normalized angular frequency.

\subsection{System Overview}%
\label{subsec:system_overview}%

\begin{figure}%
	\centering%
	\includegraphics[width=.8\linewidth]{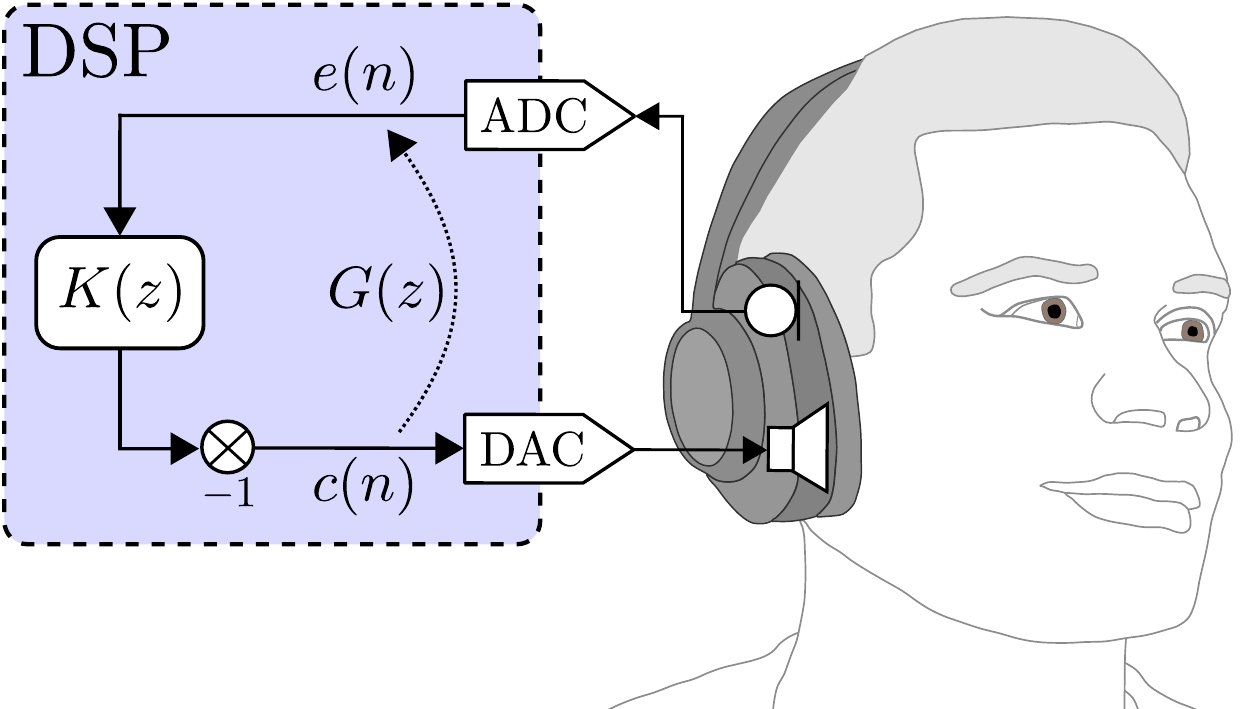}%
	\caption{Schematic layout of a digitally implemented feedback \ac{ANC} system (illustration is not to scale)}%
	\label{fig:schematic_layout}%
\end{figure}

\figref{fig:schematic_layout} shows the considered \ac{ANC} headphones, which feature a microphone and a loudspeaker that face inside the ear canal, a \ac{DSP} that implements the feedback controller $\fd{\controller}(\z)$, \acp{ADC} and \acp{DAC}. The \ac{ANC} system aims to attenuate the primary disturbance $\disturbance(\dt)$ at the microphone, which corresponds approximately to sounds at the wearer's ear drum for low frequencies.

Passive and active attenuation combined make up the total attenuation, where our focus is on the latter. The active part uses a feedback loop based on the digital error signal $\error(\dt)$, which is captured by the microphone, filtered with the impulse response of the controller $\fd{\controller}(\z)$, and inverted to yield the digital control signal $\controlsignal(\dt)$. The anti-noise signal to suppress the primary disturbance results from playing back $\controlsignal(\dt)$ over the loudspeaker, and ideally has the same amplitude as $\disturbance(\dt)$ but inverted phase.

\subsection{The Controlled System}%
\label{subsec:the_controlled_system}%

The controlled system $\fd{\plant}(\z)$, also referred to as discrete-time secondary path, is assumed to be a \ac{LTI} model. It describes the electro-acoustic system, which includes loudspeaker, microphone, \ac{ADC} and \ac{DAC}, as well as the acoustic sound transmission. In general, $\fd{\plant}(\z)$ is stable, non-minimum-phase, and can be estimated from measurements.

Knowledge about $\fd{\plant}(\z)$ is crucial for \ac{ANC} applications. The variability of $\fd{\plant}(\z)$ depends on multiple factors and is well-studied~\cite{guldenschuh14, moller95}. The shape of the wearer's head and ears are known to cause variations, but manufacturing tolerances, wear and tear, or temperature differences may also impact $\fd{\plant}(\z)$. For a single wearer, the fit of the device on or in the ears causes further variations that potentially change over time, \eg, with movements of the jaw or head. 

We divide fits into ``normal'', ``loose'' and ``tight'', and consider a fit to be normal when the device is mounted as the manufacturer intends. If the effective air volume that the headphones enclose is large compared to the normal fit, \eg, due to a leakage, we consider the fit to be loose. Conversely, a tight fit is characterized by a smaller-than-usual volume. These fits can occur, \eg, when mounting, handling, or adjusting the headphones.

\subsection{Design Specifications}%
\label{subsec:design_specifications}%

The active attenuation which the system achieves is quantified by the closed-loop response, also denoted ``sensitivity'' or ``disturbance rejection''. It is defined as
\begin{align}\label{eq:sensitivity}%
	\fd{\sensitivity}(\z) = \frac{\fd{\error}(\z)}{\fd{\disturbance}(\z)} = \frac{1}{1 + \fd{\plant}(\z) \fd{\controller}(\z)} = \frac{1}{1 + \fd{\openloop}(\z)},%
\end{align}%
where $\fd{\openloop}(\z) = \fd{\plant}(\z) \fd{\controller}(\z)$ denotes the open-loop response. Note that $\fd{\sensitivity}(\z)$ is subject to Bode's sensitivity integral, which states that attenuation of sound in one frequency range causes amplifications in another~\cite{bai97, skogestad05}.

The target $\fd{\sensitivity}(\z)$ depends on frequency and on the application. A common goal is the minimization of $\abs{\fd{\sensitivity}(\z)}$ for low to middle frequencies in normal fit wearing situations, so that \ac{ANC} supplements the passive attenuation, which is typically more effective to higher frequencies~\cite{hansen12}. 

A frequent requirement is robust stability, which is given if the closed-loop remains stable despite variations in $\fd{\plant}(\z)$. State-of-the-art designs therefore use, \eg, least-squares or infinity-norm objective functions in combination with hard constraints for robust stability~\cite{bai97,rafaely99}.

\subsection{Internal Model Control}%
\label{subsec:internal_model_control}%

\begin{figure}%
	\centering%
	\includegraphics[width=\linewidth]{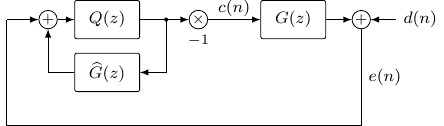}%
	\caption{System-theoretic block diagram of a feedback \ac{ANC} system using the \ac{IMC} configuration}%
	\label{fig:imc_block_diagram}%
\end{figure}

The \ac{IMC} configuration is a commonly-used concept in designing $\fd{\controller}(\z)$, which has first been proposed in~\cite{francis76}, and applied to \ac{ANC} in~\cite{rafaely99}. \figref{fig:imc_block_diagram} shows a system-theoretic block diagram of the considered \ac{IMC}-based feedback system. It consists of the controlled system $\fd{\plant}(\z)$ and the controller $\fd{\controller}(\z)$, which features a feed-forward filter $\fd{\q}(\z)$ and a fixed internal model $\nominal{\fd{\plant}}(\z)$ which is connected to $\fd{\q}(\z)$ in a feedback loop. If $\fd{\q}(\z)$ is a \ac{FIR} filter, this concept is widely regarded as \ac{FIR}-Q-parameterization~\cite{elliott00}. The transfer function $\fd{\controller}(\z)$ is then given by
\begin{align}%
	\label{eq:imc_controller}%
	\fd{\controller}(\z) = \frac{\fd{\q}(\z)}{1 - \nominal{\fd{\plant}}(\z) \fd{\q}(\z)}.
\end{align}%

The purpose of \ac{IMC} is to simplify the optimization. It guarantees nominal stability, \ie, the feedback loop remains stable for any $\fd{\q}(\z)$ if $\fd{\plant}(\z) = \nominal{\fd{\plant}}(\z)$ holds. Moreover, the feedback cancels out the denominator of the nominal sensitivity $\nominal{\fd{\sensitivity}}(\z)$, as can be seen when inserting \eqref{eq:imc_controller} into \eqref{eq:sensitivity}:
\begin{align}%
	\nominal{\fd{\sensitivity}}(\z) &= 1 - \fd{\q}(\z) \nominal{\fd{\plant}}(\z). \label{eq:imc_sensitivity}
\end{align}%
Further measures for the more rigorous robust stability requirement that exceeds $\nominal{\fd{\plant}}(\z)$ will be discussed in \secref{sec:controller_optimization}.

%% file: headphonemeasurements.tex
\section{Headphone Measurements}%
\label{sec:headphone_measurements}%

We studied the acoustic behavior of $\fd{\plant}(\z)$ for exemplary \ac{ANC} headphones through extensive measurements with human test subjects and a dummy head. The aim was to analyze the uncertainty which the headphones face in different operating conditions. Besides normal fits, which are relevant for optimizing \ac{ANC} performance, we also consider loose or tight fits, which substantially increase the amount of variations and make the guarantee for stability challenging in practice. The study focuses on over-ear headphones, but we applied the same procedure to a pair of in-ear headphones as well. This is not detailed here in order to avoid a lengthy discussion, and we refer the interested reader to~\cite{hilgemann22} for further details.

\subsection{Measurement Setup}%
\label{subsec:measurement_setup}%

We consider two devices suited to implement \ac{ANC}: a pair of Bose QC45 over-ear headphones, and a pair of Bose QC20 in-ear headphones. We removed the manufacturer's \ac{ANC} electronics, and connected a live-system with a sound card to the built-in microphones and loudspeakers. Specifically, an Analog Devices ADAU1787 audio codec which runs at a sampling rate $\samplerate=192\,\mathrm{kHz}$ was used. The effect of this codec had to be considered explicitly in the measurements because it is part of the controlled system $\fd{\plant}(\z)$.

We used a log-sweep of $10$ seconds duration as excitation~\cite{mueller01}. The parameters were chosen to yield a good \ac{SNR} without causing discomfort to the subjects. We performed all measurements in a measurement room whose reverberation time, noise rating and operational room response curve complied with the recommendation ITU-R BS.1116-2~\cite{ITU}.

\begin{figure}%
	\centering%
	\includegraphics[width=\linewidth]{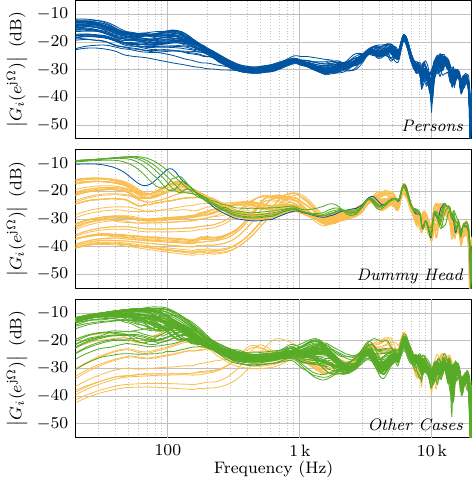}%
	\caption{Measured responses for QC45 headphones: human wearer normal fits (top), different fits on a dummy head (middle), and other cases (bottom). The colors \mbox{(\refloosefit/\refnormalfit/\reftightfit)} indicate loose, normal and tight fits, respectively.}%
	\label{fig:qc45_magnitude_responses_persons}%
\end{figure}

The measurement series consisted of three parts, with $222$ and $166$ measurements for over-ear and in-ear headphones, respectively. In the following, we consider specific observations of $\fd{\plant}(\z)$, that we denote as  $\fd{\plant}_{\observationidx}(\z)$ for $1 \leq \observationidx \leq \numobservations$, and $\numobservations$ refers to the number of measurements. We used spectral division to obtain the $\fd{\plant}_{\observationidx}(\z)$ as \ac{FIR} filters as described in~\cite{mueller01}. \figref{fig:qc45_magnitude_responses_persons} shows the left channel of the measured magnitude responses $\abs{\fd{\plant}_{\observationidx}(\onunitcircle)}$ for the over-ear headphones.

The first part of the measurement series analyzed variations of the normal fit response for $35$ different human test subjects aged between $18$ and $61$ years. We instructed the subjects to adjust the headphones according to their subjective wearing comfort. For the over-ear headphones, this included adjustments of the headband and cushions, whereas for the in-ear headphones, subjects could select one of three differently-sized silicone ear tips. People that use glasses regularly were instructed to keep them on.

Part two of the measurements involved to induce loose fits and tight fits of different severity in a reproducible way using a Head Acoustics HMS II.3 dummy head. We induced leakages with different sizes, \eg, by using a spectacle frame or by pulling the headphones away from the head using straps. Conversely, we used a strap to press the headphones against the dummy head with varying amounts of force to obtain different tight fits.

The final part considers cases that are mostly agnostic to the wearer, including the ``open'' case (headphone cans lie face-up on a table) and several ``closed'' fits (\eg, headphone cans lie face-down, or cushions manually blocked). We paid special attention to the tight fits as they produced the largest magnitude responses.

\subsection{Analysis of Variations}%
\label{subsec:analysis_of_variations}%

\figref{fig:qc45_magnitude_responses_persons} shows that the gain varies between individuals by about $3\,\mathrm{dB}$ to $5\,\mathrm{dB}$ between $300\,\mathrm{Hz}$ and $1\,\mathrm{kHz}$, and more than $10\,\mathrm{dB}$ at lower frequencies. These variations are similar to what is reported in the literature~\cite{moller95,masiero11,hiipakka10}. The dummy head measurements show a drop of the main resonance below $200\,\mathrm{Hz}$ and a shift towards higher frequencies for loose fits. This increased with the leak size: the observation with smallest magnitude at $200\,\mathrm{Hz}$ corresponds to the greatest leakage. Conversely, the tight fit magnitude response increased by more than $6\,\mathrm{dB}$ compared to the normal fit, which is visible from the dummy head and handling case measurements. The magnitude response variance depends on frequency and exceeds $30\,\mathrm{dB}$ at low frequencies.

\begin{figure}%
	\hspace{0.02\linewidth}%
	\begin{subfigure}{.31\linewidth}%
		\centering%
		\includegraphics[height=\linewidth]{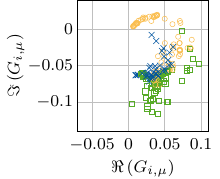}%
		\caption{$\frequency = 200\,\mathrm{Hz}$}%
		\label{subfig:complex_plane_200Hz}%
	\end{subfigure}%
	\hspace{0.045\linewidth}%
	\begin{subfigure}{.31\linewidth}%
		\centering%
		\includegraphics[height=\linewidth]{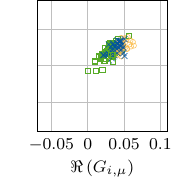}%
		\caption{$\frequency = 2.8\,\mathrm{kHz}$}%
		\label{subfig:complex_plane_2800Hz}%
	\end{subfigure}%
	\hspace{-0.02\linewidth}%
	\begin{subfigure}{.31\linewidth}%
		\centering%
		\includegraphics[height=\linewidth]{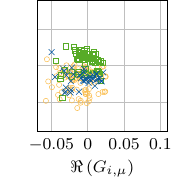}%
		\caption{$\frequency = 4.6\,\mathrm{kHz}$}%
		\label{subfig:complex_plane_4600Hz}%
	\end{subfigure}%
	\caption{Observations $\plantobservationdiscretefreq$ at different frequencies. Marks (\refscatterloose/\refscatternormal/\refscattertight) indicate loose, normal and tight fits, respectively.}%
	\label{fig:complex_plane_observations_examples}%
\end{figure}

To gain further insight into the variation of $\fd{\plant}(\onunitcircle)$, we consider the observed responses $\fd{\plant}_{\observationidx}(\onunitcircle)$ in the complex plane. We therefore evaluate $\fd{\plant}_{\observationidx}(\z)$ at specific frequency bins $\z = \onunitcirclebin$, denoted using subscript notation, \ie, $\plantobservationdiscretefreq := \fd{\plant}_{\observationidx}(\onunitcirclebin)$, where $\freqidx$ is the bin index. \figref{fig:complex_plane_observations_examples} shows the real and imaginary parts $\real{\plantobservationdiscretefreq}$ and $\imag{\plantobservationdiscretefreq}$ of the $222$ frequency responses that we measured for the QC45 headphones for exemplary frequency bins. We chose $\frequency \in \{ 200\,\mathrm{Hz}, ~2.8\,\mathrm{kHz}, ~4.6\,\mathrm{kHz} \}$ as prominent examples where the measured data shows different distributions.

The figure shows that the observations $\plantobservationdiscretefreq$ occur with different shapes. At $2.8\,\mathrm{kHz}$, the points have similar distances to the origin (\ie, magnitudes), and most variations occur along a line with about $30^\circ$ angle from the real axis. In contrast, the variations show a quasi-circular shape at $4.6\,\mathrm{kHz}$. For $200\,\mathrm{Hz}$, the $\plantobservationdiscretefreq$ are more widely dispersed and show greater variations along the imaginary axis compared to the real axis. Similar findings have been reported in our previous work on in-ear headphones~\cite{hilgemann22}, and in~\cite{rafaely99} for a car headrest system.

%% file: uncertaintymodels.tex
\section{Uncertainty Modeling}%
\label{sec:uncertainty_modeling}%

The design of feedback controllers for \ac{ANC} headphones is typically based on a model of the uncertainty that the controlled system is afflicted with. The uncertainty is often described as a set $\plantset$ of frequency responses that the controlled system assumes~\cite{skogestad05, zhou96, seiler20}. This section discusses the abstraction of observations into a model $\plantset$ which serves as basis for the design of robust controllers. In the following, we consider individual frequency bins $\freqidx$.

\begin{figure*}%
	\hspace{0.04\linewidth}
	\begin{subfigure}{.25\linewidth}%
		\centering%
		\includegraphics[height=.9\linewidth]{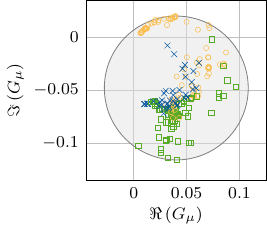}%
		\caption{Norm-bounded model}%
		\label{subfig:complex_plane_norm_bounded}%
	\end{subfigure}%
	\hspace{-0.02\linewidth}%
	\begin{subfigure}{.25\linewidth}%
		\centering%
		\includegraphics[height=.9\linewidth]{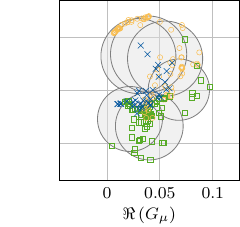}%
		\caption{Multi-disk model}%
		\label{subfig:complex_plane_multi_disk}%
	\end{subfigure}%
	\hspace{-0.04\linewidth}%
	\begin{subfigure}{.25\linewidth}%
		\centering%
		\includegraphics[height=.9\linewidth]{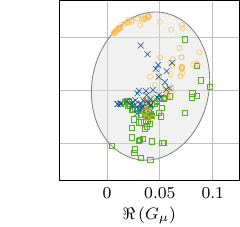}%
		\caption{Elliptic model}%
		\label{subfig:complex_plane_elliptic}%
	\end{subfigure}%
	\hspace{-0.04\linewidth}%
	\begin{subfigure}{.25\linewidth}%
		\centering%
		\includegraphics[height=.9\linewidth]{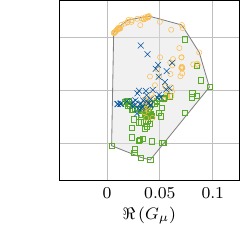}%
		\caption{Convex hull model}%
		\label{subfig:complex_plane_convex_hull}%
	\end{subfigure}%
	\caption{Models of frequency response uncertainty at $\frequency = 200\,\mathrm{Hz}$ for QC45 over-ear headphones. Colored marks \mbox{(\refscatterloose/\refscatternormal/\refscattertight)} indicate loose, normal and tight fit observations $\plantobservationdiscretefreq$, respectively. Different models $\plantsetdiscretefreq$ are shown as shaded areas.}%
	\label{fig:complex_plane_uncertainty_models}%
\end{figure*}

This section is organized as follows: we discuss general requirements on uncertainty models in~\secref{subsec:model_requirements}. We revisit conventional models in \secref{subsec:normboundedmodel} and \secref{subsec:multidiskmodel}, and propose two alternatives that aim to reduce the conservatism in \secref{subsec:ellipticmodel} and \secref{subsec:convexhullmodel}. As an example, \figref{fig:complex_plane_uncertainty_models} shows all considered models for $200\,\mathrm{Hz}$ as shaded areas. Their details will become clear as the section progresses. Finally, we compare the models in \secref{subsec:model_comparison}.

\subsection{Model Requirements}%
\label{subsec:model_requirements}%

An important aspect of uncertainty models is the introduced conservatism: $\plantsetdiscretefreq$ typically covers parts of the complex plane where no $\plantobservationdiscretefreq$ were observed~\cite{skogestad05}. This is intended to some extent, because the finite number of observations only reflect a part of the true variation. However, this also restricts the solution set when optimizing the controller, which limits the performance that can be achieved. 

The approaches studied in this work assume that the observed $\plantobservationdiscretefreq$ reliably reflect variations that occur in practice. Since the set $\plantsetdiscretefreq$ ideally consists only of $\plantobservationdiscretefreq$ that the controlled system can actually assume, we strive for a model with minimal area that retains these $\plantobservationdiscretefreq$.

We assume that the headphone fit can change, \eg, from a normal fit to a tight fit. We therefore require a contiguous model area to also cover stability when the fit changes. As an example, the tri-rectangle model~\cite{guo24} guarantees stability for the normal, loose or tight fits individually, but its area may not be contiguous. We do not consider the tri-rectangle model in the following because it does not cover stability between fits: the feedback loop can become unstable as the fit changes. In contrast, we require that the model is able to guarantee stability also between fits.

\subsection{Norm-Bounded Model}%
\label{subsec:normboundedmodel}%

The norm-bounded model is the most common uncertainty model that, \eg, \ac{IMC}-based controller designs or $\mathcal{H}_\infty$ synthesis use~\cite{bai97, rafaely99}. The set $\plantsetdiscretefreq$ is a disk with center $\circlecenterdiscretefreq{}$ and radius $\circleradiusdiscretefreq = \abs{\circlecenterdiscretefreq{} \cdot \muluncdiscretefreq{}}$, with $\muluncdiscretefreq{}$ denoting the multiplicative uncertainty bound. The corresponding set $\plantsetdiscretefreq$ is defined as
\begin{align}%
	\normbounded{\plantsetdiscretefreq} = \set{ \fd{\plant} \in \mathbb{C} ~ \Big| ~ \abs{\circlecenterdiscretefreq{} - \fd{\plant}} \leq \circleradiusdiscretefreq },%
\end{align}%
and its parameters $\circlecenterdiscretefreq{}$ and $\muluncdiscretefreq{}$ can be obtained from the well-known smallest circle problem~\cite{chrystal85}.

The main advantage of the norm-bounded model is its simplicity, which offers straightforward means for control systems analysis and synthesis. A disadvantage is that the uncertainty is overestimated when the observations $\plantobservationdiscretefreq$ are not distributed in a circular shape. We see this in \figref{subfig:complex_plane_norm_bounded}, where we did not observe any responses with a negative real part, yet this part is covered by the model.

\subsection{Multi-Disk Model}%
\label{subsec:multidiskmodel}%

In order to reduce the conservatism of the norm-bounded model, we proposed a conceptually similar model in a previous work~\cite{hilgemann22}. More precisely, we proposed to define $\plantsetdiscretefreq$ as the union $\cup$ of $\numdisks$ smaller disks:
\begin{align}%
	\multidisk{\plantsetdiscretefreq} = \bigcup_{\circleidx=1}^{\numdisks} \set{ \fd{\plant} \in \mathbb{C} ~ \Big| ~ \abs{\circlecenterdiscretefreq{\circleidx} - \fd{\plant}} \leq \circlesradiusdiscretefreq{\circleidx} }.
\end{align}%
As before, $\circlecenterdiscretefreq{\circleidx}$ and $\circlesradiusdiscretefreq{\circleidx}$ refer to the disk centers and radii. The model parameters can be obtained using the algorithm from~\cite{hilgemann22}, which uses a shared point across circles to obtain a contiguous model area.

\figref{subfig:complex_plane_multi_disk} shows the structure of the multi-disk model with $\numdisks=6$ disks. We observe that the use of smaller disks allows to define $\plantsetdiscretefreq$ in a more fine-grained manner. Compared to the norm-bounded model, the area is reduced by omitting those areas with few or no observations. The area which covers a negative real part was reduced, but it could evidently be reduced further. A disadvantage is the greater number of constraints needed to design the controller.

\subsection{Elliptic Model}%
\label{subsec:ellipticmodel}%

Since the variation of $\plantobservationdiscretefreq$ in \figref{fig:complex_plane_uncertainty_models} is more pronounced along the imaginary axis compared to the real axis, we feel that it is natural to define $\plantsetdiscretefreq$ as an ellipse
\begin{align} \label{eq:inside_of_ellipse_inequality}%
	\elliptic{\plantsetdiscretefreq} = \set{\fd{\plant} \in \mathbb{C} ~ \Bigg| ~ \left( \frac{\ellipmajorauxdiscretefreq}{\ellipmajordiscretefreq} \right)^2 + \left( \frac{\ellipminorauxdiscretefreq}{\ellipminordiscretefreq} \right)^2 \leq 1 },
\end{align}%
where we introduced auxiliary terms
\begin{equation}%
	\begin{aligned}%
		\ellipmajorauxdiscretefreq &= \cos(\ellipanglediscretefreq) \real{\Delta\plantdiscretefreq} + \sin(\ellipanglediscretefreq) \imag{\Delta\plantdiscretefreq}, \\
		\ellipminorauxdiscretefreq &= \sin(\ellipanglediscretefreq) \real{\Delta\plantdiscretefreq} - \cos(\ellipanglediscretefreq) \imag{\Delta\plantdiscretefreq},
	\end{aligned}%
\end{equation}%
and $\Delta\plantdiscretefreq = \fd{\plant} - \ellipcenterdiscretefreq$. These equations are based on the geometric form of an ellipse with center $\ellipcenterdiscretefreq$, semi-major axis $\ellipmajordiscretefreq$, semi-minor axis $\ellipminordiscretefreq$, and the angle $\ellipanglediscretefreq$ between $\ellipmajordiscretefreq$ and the real axis. The parameters can be determined from the smallest enclosing ellipse (the Löwner-John ellipsoid) that is found, \eg, using Welzl's algorithm~\cite{welzl91} or convex optimization~\cite{kumar05}. The norm-bounded model is included as a special case for $\ellipmajordiscretefreq = \ellipminordiscretefreq$.

\subsection{Convex Hull Model}%
\label{subsec:convexhullmodel}%

The model area can be further reduced by using the smallest convex area to contain all observations. This polyhedral shape, commonly referred to as convex hull, is defined as the intersection $\cap$ of $\numhalfspaces$ half-space sets:
\begin{equation}%
	\begin{aligned} \label{eq:inside_of_convex_hull_inequality}%
		\convexhull{\plantsetdiscretefreq} = \bigcap_{\halfspaceidx=1}^{\numhalfspaces} \Big\{ \fd{\plant} \in \mathbb{C} ~ \Big| ~ & \firsthalfspacevardiscretefreq \real{\fd{\plant}} + \\ &\secondhalfspacevardiscretefreq \imag{\fd{\plant}} + \halfspaceoffsetdiscretefreq \leq 0 \Big\}.
	\end{aligned}%
\end{equation}%
The model parameters are the number of half-spaces $\numhalfspaces$, the weights $\firsthalfspacevardiscretefreq$ and $\secondhalfspacevardiscretefreq$, and the offsets $\halfspaceoffsetdiscretefreq$, which can be determined using the quickhull algorithm~\cite{barber96}. An advantage of using a convex shape is that transitions between observations are captured inherently.

\subsection{Model Comparison}%
\label{subsec:model_comparison}%

\begin{figure}%
	\centering%
	\includegraphics[width=\linewidth]{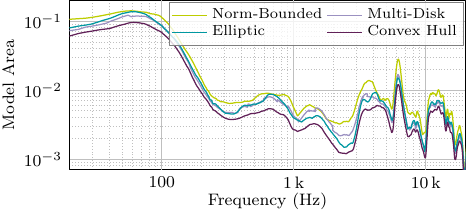}%
	\caption{Area over frequency of different uncertainty models for the Bose QC45 over-ear headphones}%
	\label{fig:area_over_frequency}%
\end{figure}

To analyze variations in $\fd{\plant}_{\observationidx}$, we derived the parameters for all previously discussed models using our measurement data. \figref{fig:area_over_frequency} compares the resulting areas over frequency for all models, where lower areas indicate less modeled uncertainty. The areas generally show the same trend: the highest variations occur at low frequencies, as can also be seen in~\figref{fig:qc45_magnitude_responses_persons}.

The norm-bounded model generally yields the largest areas while the convex hull model consistently yields the smallest area. On average, this area covers about $60\,\%$ of the area for the norm-bounded model. The conventional model is accurate when the variations show a circle-like pattern, \eg, at $4.6\,\mathrm{kHz}$, where the area of the convex hull model is $81\,\%$ as large as the area of the norm-bounded model. However, the disk overestimates the uncertainty at $2.8\,\mathrm{kHz}$ since it is possible to find a convex hull that covers only a third of the area. The multi-disk and elliptic models generally lie in between the norm-bounded and convex hull models, but show minor frequency-dependent differences.

%% file: controlleroptimization.tex
\section{Controller Optimization}%
\label{sec:controller_optimization}%

In this section, we describe the optimization of $\fd{\controller}(\z)$ for \ac{ANC} headphones based on the \ac{IMC} configuration. To guarantee robustness against model uncertainty, we derive constraint functions based on the models outlined in~\secref{sec:uncertainty_modeling}. These models are fine-tuned to the observed variations, which distinguishes our work from previous works.

\subsection{Optimization Problem}%
\label{subsec:optimization_problem}%

We describe an optimization framework similar to~\cite{rafaely99} which accommodates frequency-dependent design specifications. Therefore, we use a dense grid of $\numfreq$ bins $(1\leq\freqidx\leq\numfreq)$ and assume that $\numfreq$ is chosen large enough to accurately model the dynamics. We optimize $\fd{\controller}(\z)$ indirectly through $\fd{\q}(\z)$ using \ac{IMC} as laid out in~\secref{subsec:internal_model_control}, and define the length $\numtaps$ vector $\filtercoefficients$ to hold the impulse response $\q(\dt)$ as optimization variable:
\begin{align}%
	\filtercoefficients = \trans{%
		\begin{bmatrix}%
			\q(0) & \q(1) & q(2) & \cdots & \q(\numtaps-1)%
		\end{bmatrix}%
	}.%
\end{align}%
We define the frequency domain objective function $\loss(\filtercoefficients)$ to evaluate the average frequency-weighted nominal sensitivity over $\numfreq$ frequency bins:
\begin{align}\label{eq:loss_function}%
	\loss(\filtercoefficients) = \frac{1}{\numfreq} \sum_{\freqidx=1}^{\numfreq} \abs{\performanceweightdiscretefreq \cdot \left[1 - \fd{\plantdiscretefreq} \fd{\qdiscretefreq}(\filtercoefficients) \right]}^2.
\end{align}%
Here, $\fd{\qdiscretefreq}(\filtercoefficients)$ corresponds to bin $\freqidx$ of the \ac{DFT} of $\q(\dt)$, evaluated by the inner product $\trans{\zvec_{\freqidx}}\filtercoefficients$, where
\begin{align}%
	\zvec_{\freqidx} =%
	\trans{%
		\begin{bmatrix}%
			1 & e^{\iu\normfreqbin} & e^{2\iu\normfreqbin} & \cdots & e^{(\numtaps-1)\iu\normfreqbin}
		\end{bmatrix}%
	}.%
\end{align}%
A common choice for $\performanceweightdiscretefreq$ is the square-root of the A-weighted \ac{PSD} of $\disturbance(\dt)$, in which case the controller minimizes the variance of $\error(\dt)$~\cite{rafaely99}. 

An unconstrained minimization of \eqref{eq:loss_function} is generally not feasible as it might cause instability if $\fd{\plant}(\z)$ and $\nominal{\fd{\plant}}(\z)$ differ. Robust stability is achieved if the set of open-loop responses excludes the critical point for all $\normfreq$~\cite{skogestad05}. To achieve this, we impose $\numfreq$ hard constraints $\constraintdiscretefreq(\filtercoefficients)$ on $\filtercoefficients$. Each $\constraintdiscretefreq(\filtercoefficients)$ is a scalar function which evaluates to $\constraintdiscretefreq(\filtercoefficients) < 0$ if the candidate filter $\filtercoefficients$ satisfies the robust stability criterion at frequency bin $\freqidx$, and $\constraintdiscretefreq(\filtercoefficients) \geq 0$ otherwise. The resulting problem is a nonlinear optimization problem with nonlinear constraints
\begin{equation}%
	\begin{aligned}%
		\label{eq:optimization_problem}%
		& \underset{\filtercoefficients}{\text{minimize}} \quad \loss(\filtercoefficients) \\
		& \text{subject to} \quad \constraintdiscretefreq(\filtercoefficients) < 0,  \quad 1 \leq \freqidx \leq \numfreq,%
	\end{aligned}%
\end{equation}%
Note that \eqref{eq:optimization_problem} is a convex optimization problem with a guaranteed global optimum when $\constraintdiscretefreq(\filtercoefficients)$ is based on the norm-bounded or multi-disk models. For its specific definition, we refer to~\cite{rafaely99} and~\cite{hilgemann22}. In the following, we develop $\constraintdiscretefreq(\filtercoefficients)$ for the elliptic and convex hull models.

\subsection{Novel Constraints}%
\label{subsec:novel_constraints}%

\begin{figure}%
	\centering%
	\includegraphics[width=.7\linewidth]{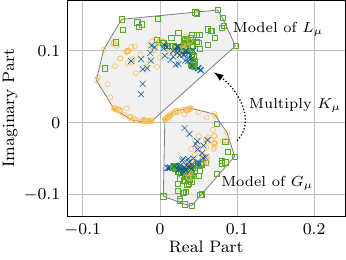}%
	\caption{Relationship between controlled system and open-loop uncertainty models for $\frequency=200\,\mathrm{Hz}$}%
	\label{subfig:open_loop_visualization}%
\end{figure}

For robust stability, the critical point needs to be excluded from the set of open-loop responses. It is therefore instructive to understand how this set is ``generated'': for all bins $\freqidx$, the $\openloopdiscretefreq$ result from the multiplication $\controllerdiscretefreq \cdot \plantdiscretefreq$. This linear operation corresponds to a multiplication of their absolute values and the summation of their angles: 
\begin{equation}%
	\begin{aligned}%
		\abs{\openloopdiscretefreq} &= \abs{\controllerdiscretefreq} \cdot \abs{\plantdiscretefreq}, \\
		\angle \openloopdiscretefreq &= \angle \controllerdiscretefreq + \angle \plantdiscretefreq.
	\end{aligned}%
\end{equation}%
The resulting set of open-loop responses at bin $\freqidx$ thus corresponds to $\plantsetdiscretefreq$ with scaling and rotation. \figref{subfig:open_loop_visualization} visualizes this for $\frequency = 200\,\mathrm{Hz}$ using $\controllerdiscretefreq = -1+2\iu$ and the convex hull model from \figref{subfig:complex_plane_convex_hull} as an example. The robust stability criterion is satisfied for the chosen $\freqidx$ because the critical point is not in the set of open-loop responses. However, a significant increase in $\abs{\controllerdiscretefreq}$ would expand the area to also cover the critical point.

The key is to formalize the effect of $\controllerdiscretefreq$ on the model parameters, which allows to write explicit constraint functions that are suitable for controller optimization. In the following, we show how the robust stability constraints can be formalized as a function of the optimization variable $\filtercoefficients$.

\subsubsection{Elliptic Constraint}%
\label{subsubsec:elliptic_constraint}%

If we multiply all points within $\elliptic{\plantsetdiscretefreq}$ by $\controllerdiscretefreq$, we obtain a new ellipse with different model parameters
\begin{equation} \label{eq:new_ellipse}%
	\begin{aligned}%
		\modified{\ellipmajordiscretefreq} &= \abs{\controllerdiscretefreq}\ellipmajordiscretefreq, &  
		\modified{\ellipminordiscretefreq} &= \abs{\controllerdiscretefreq}\ellipminordiscretefreq, \\
		\modified{\ellipanglediscretefreq} &= \ellipanglediscretefreq + \angle \controllerdiscretefreq, & 
		\modifiedellipcenterdiscretefreq &= \controllerdiscretefreq \ellipcenterdiscretefreq.
	\end{aligned}%
\end{equation}%
To derive $\elliptic{\constraintdiscretefreq}(\filtercoefficients)$, we use the inequality that defines the ellipse in \eqref{eq:inside_of_ellipse_inequality}, but using the parameters given by \eqref{eq:new_ellipse}. In the resulting equation, we replace the ``$\leq$'' by a ``$>$'' to test if a given test point is on the outside rather than on the inside. Finally, we use the critical point as test point and substitute $\controllerdiscretefreq$ according to \eqref{eq:imc_controller} to get an expression that depends on $\filtercoefficients$. Elementary rearrangements and simplifications lead to the constraint function
\begin{align}%
	\label{eq:elliptical_constraint_with_IMC}
	\elliptic{\constraintdiscretefreq}(\filtercoefficients) = \abs{\qdiscretefreq(\filtercoefficients)}^4 - \frac{\modified{\ellipmajorauxdiscretefreq}^2(\filtercoefficients)}{\ellipmajordiscretefreq^2} - \frac{\modified{\ellipminorauxdiscretefreq}^2(\filtercoefficients)}{\ellipminordiscretefreq^2},
\end{align}%
where we defined
\begin{equation}%
	\begin{aligned}%
		\modified{\ellipmajorauxdiscretefreq}(\filtercoefficients) &= \cos(\ellipanglediscretefreq) \real{\ellipaux(\filtercoefficients)} - \sin(\ellipanglediscretefreq) \imag{\ellipaux(\filtercoefficients)}, \\
		\modified{\ellipminorauxdiscretefreq}(\filtercoefficients) &= \sin(\ellipanglediscretefreq) \real{\ellipaux(\filtercoefficients)} + \cos(\ellipanglediscretefreq) \imag{\ellipaux(\filtercoefficients)},
	\end{aligned}%
\end{equation}%
and
\begin{align}
	\label{eq:elliptic_auxiliary_with_IMC}
	\ellipaux(\filtercoefficients) = \qdiscretefreq(\filtercoefficients) + \abs{\qdiscretefreq(\filtercoefficients)}^2 \conj{(\ellipcenterdiscretefreq - \nominal{\fd{\plant}}_\freqidx)}.
\end{align}
Here, an asterisk ($\conj{}$) denotes the conjugate of a complex number. Note that \eqref{eq:elliptical_constraint_with_IMC} is generally a non-convex function of $\filtercoefficients$ since $\modified{\ellipmajorauxdiscretefreq}(\filtercoefficients)$ and $\modified{\ellipminorauxdiscretefreq}(\filtercoefficients)$ are subtracted.

\subsubsection{Convex Hull Constraint}%
\label{subsubsec:convex_hull_constraint}%

The multiplication of all elements in $\convexhull{\plantsetdiscretefreq}$ by $\controllerdiscretefreq$ yields a different convex hull with modified parameters
\begin{equation}%
	\begin{aligned}%
		\modified{\convexhullanglediscretefreq{\halfspaceidx}} &= \convexhullanglediscretefreq{\halfspaceidx} + \angle \controllerdiscretefreq, \\
		\modified{\halfspaceoffsetdiscretefreq} &= \abs{\controllerdiscretefreq} \halfspaceoffsetdiscretefreq,
	\end{aligned}%
\end{equation}%
where $\convexhullanglediscretefreq{\halfspaceidx}$ denotes the counterclockwise angle between the $\halfspaceidx$-th half-space and the real axis, and $\halfspaceoffsetdiscretefreq$ is the offset as before. \figref{subfig:open_loop_visualization} visualizes this for an exemplary frequency bin $\freqidx$ which corresponds to $\frequency=200\,\mathrm{Hz}$.

Equation~\eqref{eq:inside_of_convex_hull_inequality} shows that a test point is inside the convex hull if $\numhalfspaces$ inequalities hold true. Conversely, at least one of the inequalities must be violated if a point is to lie outside. We therefore replace the ``$\leq$'' of the inequality in \eqref{eq:inside_of_convex_hull_inequality} by a ``$>$'', and compute the minimum of the resulting $\numhalfspaces$ expressions. As before, we use $-1$ as test point and substitute $\controllerdiscretefreq$ to obtain a constraint function that depends on $\filtercoefficients$:
\begin{align}%
	\label{eq:convex_hull_constraint_with_IMC}%
	\convexhull{\constraintdiscretefreq}(\filtercoefficients) = \min\left( \convexhullauxfirst{1}(\filtercoefficients), ~ \ldots, ~ \convexhullauxfirst{\numhalfspaces}(\filtercoefficients) \right)
\end{align}%
with auxiliary variables
\begin{equation}%
	\begin{aligned}%
		\convexhullauxfirst{\halfspaceidx}(\filtercoefficients) &= \cos(\convexhullanglediscretefreq{\halfspaceidx}) \real{\convexhullauxsecond(\filtercoefficients)} \\ &- \sin(\convexhullanglediscretefreq{\halfspaceidx}) \imag{\convexhullauxsecond(\filtercoefficients)} - \abs{\qdiscretefreq(\filtercoefficients)}^2\halfspaceoffsetdiscretefreq, \\
		\convexhullauxsecond(\filtercoefficients) &= \qdiscretefreq(\filtercoefficients) - \abs{\qdiscretefreq(\filtercoefficients)}^2 \conj{\nominal{\fd{\plant}}_\freqidx}.
	\end{aligned}%
\end{equation}%
Equation~\eqref{eq:convex_hull_constraint_with_IMC} uses the non-smooth $\min$-function. To circumvent problems that this non-smoothness might cause in a gradient-based optimization, the approximation $\min(x_1,~\ldots,~x_{\numhalfspaces}) \approx -\frac{1}{\minapproxvar} \log \sum_{\halfspaceidx=1}^{\numhalfspaces}\exp(-\minapproxvar x_{\halfspaceidx})$ can be used. This smooth function approaches the $\min$-function for $\minapproxvar \rightarrow \infty$~\cite{bullen03}. Note that~\eqref{eq:convex_hull_constraint_with_IMC} is a non-convex function of $\filtercoefficients$.

%% file: case_study.tex
\section{Case Study}%
\label{sec:case_study}%

We implemented real-time prototype feedback \ac{ANC} systems based on the four uncertainty models presented in \secref{sec:uncertainty_modeling} in order to compare their implications for a realistic use-case using the hardware described in~\secref{sec:headphone_measurements}. The goal was to maximize the active attenuation at lower frequencies while guaranteeing robust stability. The widely-used norm-bounded model and the recently proposed multi-disk model~\cite{hilgemann22} have been parametrized to get the best possible performance, and served as baseline methods.

\subsection{System Design}%
\label{subsec:system_design}%

We used the following optimization parameters: we set $\numtaps=8192$ and used a frequency discretization with $\numfreq=8192$ linearly spaced bins in the frequency range $0\,\mathrm{Hz} < \frequency < 24\,\mathrm{kHz}$. The average of $35$ normal fits for different wearers served as internal model $\nominal{\fd{\plant}}(\z)$. To put emphasis on the attenuation at lower frequencies, we specified the design goal $\fd{\performanceweight}(\z)$ as an eight-order Butterworth bandpass filter with a peak gain of $31\,\mathrm{dB}$, and $0\,\mathrm{dB}$-crossovers at about $40\,\mathrm{Hz}$ and $1\,\mathrm{kHz}$. This unusually high peak gain was selected to push the limits of achievable attenuation with a robustly stable system.

We solved \eqref{eq:optimization_problem} with these parameters to optimize the controllers subject to the uncertainty models from \secref{sec:uncertainty_modeling} using an interior-point algorithm~\cite{nocedal06}. Since \eqref{eq:optimization_problem} is a convex problem for the norm-bounded and multi-disk models, we obtained their global optima. In contrast, the proposed elliptic and convex hull constraints are non-convex, so we could only find local optima. While the controller based on the norm-bounded model achieved $\normbounded{\loss}(\filtercoefficients) = 1.11$, we obtained $\multidisk{\loss}(\filtercoefficients) = 0.66$ for the multi-disk model, $\elliptic{\loss}(\filtercoefficients) = 0.56$ for the elliptic model, and $\convexhull{\loss}(\filtercoefficients) = 0.54$ for the convex hull model. This shows that the added flexibility was utilized to improve the performance.

Finally, we implemented the controllers on the \ac{DSP} hardware, which required to approximate $\fd{\controller}(\z)$ by low-order \ac{IIR} filters. We obtained $50$-th order \ac{IIR} filters using the balanced truncation algorithm described in~\cite{zhou95}. These filters deviated minimally from the high-order ones, and maintained all specifications.

\subsection{Measurement Procedure}%
\label{subsec:measurement_procedure}%

Prior to the study, we tested all controllers thoroughly for stability, \eg, by inducing different amounts of leakage on the dummy head, or by manually blocking the ear cushions or ear molds, respectively. We were not able to induce instabilities, neither with fixed loose or tight fits, nor when transitioning from one fit to another. We observed the same stability in practice for all four uncertainty models.

We proceeded to measure the \ac{ANC} performance for a total of $21$ different human subjects with ages in between $21$ and $60$. We instructed the subjects to adjust the headphones as for the measurements in \secref{subsec:measurement_setup} and to keep their heads still during the measurements. We played back diffuse pink noise over eight Neumann KH120D loudspeakers with a semi-circular, lateral placement around the listener, and one Velodyne DD-12 subwoofer to synthesize the primary noise field. The noise sound level was calibrated using a Brüel \& Kj\ae r sound level meter 2240 to obtain noise with roughly $80\,\mathrm{dBA}$ at the listener's head position.

We focus on the effect of active cancellation by comparing the measured signals recorded by the built-in microphones with \ac{ANC} switched off and on, respectively. For each test subject, we performed one calibration measurement with no controller active, and one with either of the controllers active, with a measurement duration of $10\,\mathrm{s}$. For this, we reconfigured the \ac{DSP} to use one of four controllers on the fly in a randomized order. The headphones were not re-fitted between these measurements to capture the same variations across different models. To estimate only the active part of the measured total attenuation, we spectrally divided these measurements by the calibration data, which removes the effect of passive attenuation. 

Although not relevant for the user's perception of \ac{ANC} performance, we also measured the performance with loose and tight fits to confirm the stability in challenging wearing situations. We used the dummy head for these measurements, which allowed to induce different fits as in \secref{sec:headphone_measurements} in a reproducible manner.

\subsection{Measurement Results}%
\label{subsec:performance_measurement_results}%

\begin{figure*}%
	\centering%
	\includegraphics[width=\textwidth]{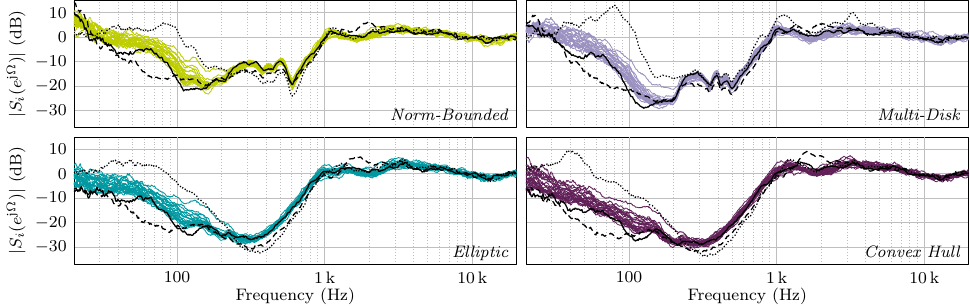}%
	\caption{Left channel closed-loop response measured for the QC45 headphones with $1/6$-th octave band smoothing. Colored lines (\refnormbounded/\refcenteredmultidisk/\refelliptic/\refconvexhull) indicate normal fits of $21$ persons for the norm-bounded, multi-disk, elliptic, and convex-hull models, respectively. Black lines (\refdhloose/\refdhnormal/\refdhtight) indicate loose, normal and tight fits, respectively, using a dummy head.}%
	\label{fig:qc45_performance_measurement_all}%
\end{figure*}

\figref{fig:qc45_performance_measurement_all} shows the measured closed-loop magnitude responses $\abs{\fd{\sensitivity}_{\observationidx}(\onunitcircle)}$ for the left channel of $21$ human wearers (thin lines) and for the dummy head (thick lines). First, we focus on the \ac{ANC} performance for the human subjects.

The norm-bounded model results in $10\,\mathrm{dB}$ to $20\,\mathrm{dB}$ attenuation between $100\,\mathrm{Hz}$ and $800\,\mathrm{Hz}$, with a narrow-band peak of about $22\,\mathrm{dB}$ at $150\,\mathrm{Hz}$ and at $700\,\mathrm{Hz}$. A lower peak attenuation is obtained compared to other models, but there is also less waterbed overshoot. In comparison, the multi-disk model results in a slightly higher attenuation below $800\,\mathrm{Hz}$, with a peak of $29\,\mathrm{dB}$ at $140\,\mathrm{Hz}$ for one subject. Both results deviate from the design goal between $200\,\mathrm{Hz}$ and $500\,\mathrm{Hz}$ due to the restrictiveness of the respective constraints. At these frequencies, the proposed elliptic and convex hull models lead to a significant improvement, with a peak attenuation of $29\,\mathrm{dB}$ at $280\,\mathrm{Hz}$, and $31\,\mathrm{dB}$ at $240\,\mathrm{Hz}$, respectively. Below $200\,\mathrm{Hz}$, the slope of the curves is comparable for the elliptic and convex hull models, and less steep compared to the design goal.

\figref{fig:qc45_performance_measurement_all} also indicates variations of the \ac{ANC} performance across wearers. Comparing \figref{fig:qc45_performance_measurement_all} to \figref{fig:qc45_magnitude_responses_persons} shows that variations occurred in a similar fashion: for most persons, they have the same shape with a minor gain difference of about $3\,\mathrm{dB}$ to $5\,\mathrm{dB}$ in between $200\,\mathrm{Hz}$ and $1\,\mathrm{kHz}$, but the main resonances in $\fd{\plant}(\z)$ below $200\,\mathrm{Hz}$ caused variations of more than $10\,\mathrm{dB}$. Since $\fd{\plant}(\z)$ varied from person to person but $\fd{\controller}(\z)$ was fixed, the performance has been very similar between $200\,\mathrm{Hz}$ and $1\,\mathrm{kHz}$, and mostly similar for frequencies below $200\,\mathrm{Hz}$ across the uncertainty models. 

Additionally, \figref{fig:qc45_performance_measurement_all} shows the performance with loose, normal and tight fits using the dummy-head. This shows that the system remained stable, even in situations that are more challenging than the intended wearing situation. We observe that the attenuation with a normal fit is overall quite similar for dummy head and human subjects. Below $200\,\mathrm{Hz}$, the performance with human wearers roughly lies in between the loose and tight fit performances with the dummy head.

\subsection{Comparison of Average Performance}%
\label{subsec:comparison_of_average_performance}%

\begin{figure}%
	\centering%
	\includegraphics[width=\linewidth]{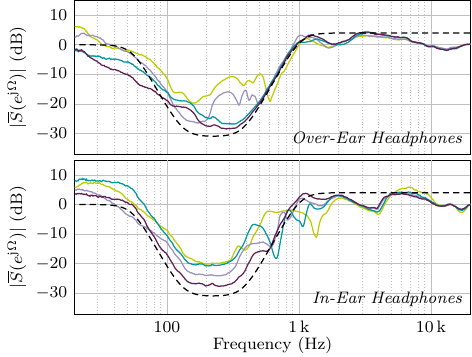}%
	\caption{Uncertainty model-dependent \ac{ANC} performance, decibel-averaged for $21$ persons with $1/6$-th octave band smoothing. Dashed lines (\reftarget) refer to the design target and solid lines (\refnormbounded/\refcenteredmultidisk/\refelliptic/\refconvexhull) to the norm-bounded, multi-disk, elliptic, and convex-hull models, respectively.}%
	\label{fig:performance_measurement_mean}%
\end{figure}

In the following, we consider the average closed-loop responses $\average{\fd{\sensitivity}}(\onunitcircle)$ for different models, which result from averaging the decibel-values of the $21$ measurements $\fd{\sensitivity}_{\observationidx}(\onunitcircle)$. \figref{fig:performance_measurement_mean} shows the resulting closed-loop responses at the left channel, together with the design goal $|\fd{\performanceweight}(\onunitcircle)|^{-1}$ for both headphones. 

We observe that the convex hull model leads to the least overall deviation from the desired behavior, followed by the elliptic, multi-disk and norm-bounded models, respectively. For the over-ear headphones, the improvement is most notable at $300\,\mathrm{Hz}$. There, the norm-bounded model yields about $11\,\mathrm{dB}$, and the convex hull model about $29\,\mathrm{dB}$ on average, which corresponds to an improvement of $18\,\mathrm{dB}$. In contrast, the norm-bounded and multi-disk models more closely reflect the design target for $30\,\mathrm{Hz}$ to $70\,\mathrm{Hz}$. The figure also indicates unwanted amplifications due to the waterbed effect. With all uncertainty models, this overshoot is distributed evenly for frequencies greater than $1\,\mathrm{kHz}$, but for both proposed models, the improved attenuation at lower frequencies comes at the cost of a more severe waterbed amplification at higher frequencies.

Using the convex hull model, the design goal is achieved more closely below $100\,\mathrm{Hz}$ for the QC20 in-ear headphones compared to the QC45 over-ear headphones, but the opposite holds true in between $300\,\mathrm{Hz}$ and $500\,\mathrm{Hz}$. Notably, the elliptic model did not yield an improvement for the in-ears, as opposed to the over-ears. Instead, the obtained performance is between the norm-bounded and multi-disk models. Both observations are due to a different manifestation of the uncertainty between in-ear and over-ear headphones, and can be explained by different model areas over frequency, as exemplified for the over-ear headphones in~\secref{subsec:model_comparison}.%  Since their model areas are similar, this underlines that the area is merely a performance indicator, but not necessarily a guarantee.

%% file: conclusion.tex
\section{Conclusion}%
\label{sec:conclusion}%

This work discussed the optimization of feedback controllers for over-ear and in-ear \ac{ANC} headphones. A robust design requires a model of the inherent uncertainty, which we derived from frequency response measurements in different wearing situations. We showed that conventional uncertainty models are more conservative than necessary to guarantee stability in practice. This limits \ac{ANC} performance since a too restrictive design constraint is implied. This work introduced uncertainty models with elliptic and polyhedral geometry to better reflect the measurements. We showed how to use these models in the \ac{IMC}-based optimization of feedback controllers.

We demonstrated the applicability of the presented models in a case study involving real-time feedback \ac{ANC} systems for two headphone types. The study compared the uncertainty models using \ac{ANC} performance measurements for an artificial dummy head and human wearers. All models showed the same robustness in practice, \ie, we did not observe any instability for either of the models. This confirms that the theoretical considerations yield a degree of stability that is sufficient in practice. Depending on frequency, the reduced conservatism allowed to increase the active attenuation below $1\,\mathrm{kHz}$ by $10\,\mathrm{dB}$ to $18\,\mathrm{dB}$ over the conventional norm-bounded uncertainty modeling approach.